\documentclass[9pt,twocolumn]{extarticle}
\usepackage[T1]{fontenc}
\usepackage[utf8]{inputenc}
\usepackage{amssymb}
\usepackage{mathptmx}
\usepackage{microtype}
\usepackage[margin=0.75in,columnsep=0.28in]{geometry}
\usepackage{amsmath}
\usepackage{booktabs}
\usepackage{graphicx}
\usepackage{tikz}
\usepackage{pgfplots}
\pgfplotsset{compat=1.17}
\usetikzlibrary{arrows.meta,positioning,fit,backgrounds,calc,shapes.multipart,shapes.geometric}
\tikzset{
  res/.style={draw=chier, line width=0.8pt, rounded corners=3pt, align=center,
              inner sep=6pt, font=\footnotesize, fill=chier!12, minimum height=9mm},
  disk/.style={draw=orange!70!black, line width=0.8pt, rounded corners=3pt, align=center,
              inner sep=6pt, font=\footnotesize, fill=orange!15, minimum height=9mm},
  op/.style={draw=black!45, rounded corners=6pt, align=center, inner sep=5pt,
             font=\footnotesize\itshape, fill=black!5, minimum height=9mm},
  plain/.style={draw=black!55, rounded corners=3pt, align=center, inner sep=6pt,
             font=\footnotesize, fill=white, minimum height=9mm},
  arr/.style={-{Stealth[length=2.2mm, width=1.8mm]}, line width=0.7pt, draw=black!65},
  darr/.style={-{Stealth[length=2.2mm, width=1.8mm]}, line width=0.7pt, draw=black!55, dashed},
  dlab/.style={font=\scriptsize\itshape, fill=white, inner sep=1.5pt},
  zone/.style={draw=black!30, dashed, rounded corners=5pt, inner sep=7pt},
  zlab/.style={font=\scriptsize\itshape\bfseries, fill=white, inner sep=2pt},
}
\usepackage[font=footnotesize,labelfont=bf]{caption}
\usepackage{titlesec}
\usepackage{balance}
\usepackage{enumitem}

\titleformat{\section}{\large\bfseries}{\thesection}{0.6em}{}
\titleformat{\subsection}{\normalsize\bfseries}{\thesubsection}{0.6em}{}
\titlespacing{\section}{0pt}{8pt plus 2pt}{4pt}
\titlespacing{\subsection}{0pt}{6pt plus 2pt}{3pt}

\definecolor{cflat}{RGB}{198,210,222}
\definecolor{cflatmt}{RGB}{130,160,190}
\definecolor{chier}{RGB}{28,60,100}
\definecolor{ccache}{RGB}{110,70,130}

\title{\bfseries\LARGE Hierarchical BM25: Lexical Search at Billion-Document Scale}
\author{Umesh Deshpande, Swaminathan Sundararaman\\[2pt] \normalsize IBM Research, San Jose, USA}
\date{}

\begin{document}

\twocolumn[
\maketitle
\vspace{-2.5em}
]

\begin{abstract}
\noindent
A flat BM25 index over one billion documents occupies about 400\,GB. Holding it in memory requires DRAM proportional to corpus size. Serving it from disk takes 4--12 seconds per query. Exact top-$k$ lexical retrieval at this scale is therefore impractical within an interactive latency budget.

\emph{Hierarchical BM25} gives up exact ranking in exchange for fixed bounds on memory and latency. A resident coarse index selects which of ${\sim}$1K topical, size-balanced document groups a query visits, using two signals: the total frequency of each query term within a group, and, for informative terms spread too thinly across groups for frequency totals to reflect, whether several of them appear together in one document. Selected groups are then searched exhaustively and scored against ${\sim}100$\,KB of global statistics. Every returned score therefore equals the flat index's score, and the approximation is confined to selection alone. The resident footprint is ${\sim}4.4$\,GB, independent of corpus size. Sixteen-term queries over one billion documents return in ${\sim}300$\,ms (4.7--5.6$\times$ the throughput of a flat multi-threaded index), and a warmed cache sustains ${\sim}32$ queries per second versus under 3 for flat indexing. At a 500K-document configuration, visiting 5--10\% of clusters recovers 0.83--0.92 of the exhaustive result score. Billion-scale recall and a direct comparison against document-reordered BlockMax-WAND remain open.

\end{abstract}

\section{Introduction}

Retrieval over large corpora uses two complementary paradigms. Lexical retrieval, anchored by BM25 within the probabilistic relevance framework~\cite{robertson2009}, scores documents from term-frequency and inverse-document-frequency statistics. It excels at exact keyword matches. Dense retrieval encodes queries and documents into a shared embedding space and ranks by vector similarity~\cite{karpukhin2020}. It stays tractable through approximate nearest-neighbor (ANN) indexes such as HNSW~\cite{malkov2018} or IVF with product quantization~\cite{jegou2011}. Neither paradigm wins on quality alone~\cite{thakur2021}. Hybrid retrieval, which runs both and merges results, is now standard practice.

This paper is about the lexical half of that pipeline---specifically, the point at which it stops scaling. A flat BM25 inverted index over one billion documents is ${\sim}400$\,GB. Keeping it in memory is uneconomical. Serving it from disk turns every multi-term query into a large fan-out of random reads, so tail latency drifts past the one-second bar interactive retrieval requires. The semantic side of hybrid retrieval is comparatively solved: mature ANN libraries scale to billions of vectors. The lexical index is the bottleneck we address.

\subsection{The trade this paper makes}
\label{sec:trade}

The design rests on a deliberate trade, stated plainly before the mechanism. \emph{Rank safety} is the following guarantee: if the true ten best documents for a query are $d_1, \ldots, d_{10}$, a rank-safe method returns exactly those ten, every time. An approximate method might return nine of them plus the eleventh-best. Dynamic-pruning methods like BlockMax-WAND are rank-safe; Hierarchical BM25 is not.

That guarantee is valuable, but consider what it costs and who needs it. In a retrieval-augmented pipeline, the lexical top-$k$ is not the final answer. It is merged with dense-retrieval candidates and passed downstream---often through a reranker---before anything reaches the user. Swapping the tenth-best lexical candidate for the eleventh rarely changes what the pipeline produces. A lexical query that takes 4--12 seconds, however, breaks the pipeline outright. An exact answer that cannot arrive inside the latency budget delivers no retrieval quality at all: the quality of a retrieval system, end to end, is bounded by whether it answers in time.

We therefore give up rank safety and buy, with it, two properties otherwise unavailable at this scale: a resident footprint fixed at ${\sim}4.4$\,GB, and query latency near 300\,ms. Both are structural bounds---consequences of the index shape---not averages that degrade under load. The sacrifice is also confined: as Section~\ref{sec:level1} shows, the approximation lives entirely in \emph{which document groups a query visits}. Every document that is scored receives its exact, corpus-wide BM25 score. The only possible failure is a missed group, never a mis-scored document. Sections~\ref{sec:quality} and~\ref{sec:discussion} state the price directly: at a 500K-document configuration the approximation recovers 0.83--0.92 of the exhaustive result score when visiting 5--10\% of clusters, while billion-scale recall remains extrapolated---so the trade is priced at small scale and stated as a bet, with analytical support, at large scale.

This paper makes three contributions, each responding to a specific gap in how lexical search is currently scaled.

\textbf{First, Hierarchical BM25 itself (Section~\ref{sec:hbm25}).} It is a two-level lexical index that decouples resident memory from corpus size. A coarse Level-1 index over ${\sim}1$K document groups is pre-computed and kept resident. A fine Level-2 index over the full billion documents is served from a fixed-size cache plus NVMe. This holds 16-term query latency at ${\sim}300$\,ms. What distinguishes this from prior two-level methods is what the first level is \emph{for}. WAND~\cite{broder2003} and recent dynamic-pruning schemes such as ASC~\cite{qiao2024} and BMP~\cite{mallia2024} assume a single index is already resident or on fast local storage; their first level skips computation \emph{within} it. At a billion documents, keeping the index resident is possible on large-memory nodes but costs DRAM proportional to corpus size. Our first level therefore bounds the \emph{residency budget}, not the computation. The architecture itself---topical shards plus per-query shard selection---is shared with selective search~\cite{kulkarni2015}; what it adds is a selection signal that sees same-document co-occurrence exactly and for every document---evidence existing shard selectors either cannot represent or observe only through a thin document sample (Section~\ref{sec:related}).

\textbf{Second, a correctness fix for cross-cluster score merging.} Building the two-level index surfaced a bug in how scores from different clusters get merged. Scoring each of ${\sim}1$K independent per-cluster indexes with its own local IDF is not an approximation---it is an outright error. Two documents with identical term frequencies can land $1.6\times$ apart in score purely because of which cluster they sit in. We fix this (Section~\ref{sec:level1}) by scoring every cluster from one ${\sim}100$\,KB table of global corpus statistics. The merged ranking then needs no per-cluster correction factor at all.

\textbf{Third, an analytical comparison against BlockMax-WAND at long query lengths (Section~\ref{sec:bmw}).} Retrieval-augmented pipelines issue much longer queries---16 to 32 terms---than the 2--5-term web queries WAND-style pruning was designed around. We argue this regime favors cluster selection. WAND's pruning power depends on document-level multi-term co-occurrence, which grows scarce as queries lengthen and terms decorrelate. Cluster selection here rests on two signals: aggregate term frequency, which needs only single-term topical concentration, a substantially weaker requirement; and exact document-level co-occurrence tracking for informative terms spread too thinly across clusters to raise any one cluster's total. For such terms, several of them meeting in one document is the evidence that marks a genuine match. We analyze why each signal's cost stays bounded as queries lengthen while WAND's does not. All of this is an architectural argument, quantified where we can, not a measured head-to-head result.

We evaluate Hierarchical BM25's latency and throughput directly at billion-document scale (Section~\ref{sec:eval}). A companion paper addresses lexical recall for the semantic vocabulary gap via query expansion. That is a distinct problem, evaluated separately.

\section{Background and Related Work}
\label{sec:related}
\label{sec:positioning}

BM25 scores a document $D$ for a query $Q$ as a sum over query terms: each term contributes an inverse-document-frequency weight times a saturated, length-normalized term-frequency factor~\cite{robertson2009}. The parameter $k_1$ caps repetition; $b$ discounts long documents. Computing the scores requires two corpus-wide statistics per term---document frequency and per-document term frequency---materialized in an inverted index; Okapi first deployed the design at TREC~\cite{robertson1994}.

Dense retrieval encodes queries and documents into a shared embedding space and ranks by vector similarity~\cite{karpukhin2020}; HNSW~\cite{malkov2018} and IVF with product quantization~\cite{jegou2011} make it tractable at billions of vectors. The semantic half of hybrid retrieval scales this way. The lexical half does not.

The baseline strategy for a multi-term BM25 query is \emph{ranked-OR}: treat the query as a disjunction, score every document matching at least one term, keep the top-$k$. Turtle and Flood analyze its two exhaustive executions---term-at-a-time and document-at-a-time~\cite{turtle1995}. Its cost is the size of the posting-list \emph{union}: ranked-OR grades every exam that answers at least one of the query's questions, and with 32 questions nearly the whole school hands one in. On this corpus the union grows from 39M to 148M documents as queries grow from 8 to 32 terms (Section~\ref{sec:wandq}), and the flat baselines of Section~\ref{sec:eval} are exhaustive ranked-OR over a disk-resident index.

Dynamic pruning accelerates ranked-OR without changing what it returns. MaxScore~\cite{turtle1995} and WAND~\cite{broder2003} maintain per-term upper bounds on score contributions and skip documents that provably cannot enter the top-$k$---a high-jump qualifier: once early jumpers set a high bar, later competitors whose personal bests fall short are waved off without jumping. BlockMax-WAND stores a bound per ${\sim}$128-posting block instead of one per term, enabling far larger skips~\cite{ding2011}. MaxScore partitions terms into essential and non-essential lists rather than pivoting through a dense document-id stream, which makes it the sturdier of the two on long disjunctive queries---a distinction Section~\ref{sec:wandq} returns to. ASC~\cite{qiao2024} and Block-Max Pruning~\cite{mallia2024} extend the idea to learned sparse retrieval. All of these are exact: they return ranked-OR's top-$k$ and differ only in how much of the union they can prove skippable.

Organized by their relation to ranked-OR: exhaustive evaluation pays for the full union; the pruning line keeps ranked-OR's semantics and shrinks its \emph{work}; Hierarchical BM25 shrinks its \emph{scope}, running plain exhaustive ranked-OR inside each visited cluster with only ${\sim}4\%$ of the corpus in the union. The pruning line also assumes an index available for traversal---resident, or on fast local storage---so its first question is how many postings a query touches. At $10^9$ documents the first question is DRAM. A flat index is ${\sim}400$\,GB: holdable on a large-memory server, but a cost that is linear in the corpus (10B documents would demand ${\sim}4$\,TB) and that competes with the dense-retrieval structures sharing the node in a hybrid deployment. The coarse level of Section~\ref{sec:hbm25} converts that linear DRAM cost into a ${\sim}4.4$\,GB constant (Table~\ref{tab:memory}). The two approaches compose: BMW-style pruning can run inside each selected cluster---though unlike BMW, cluster selection here is not rank-safe.

A complementary line reorders document ids instead of pruning: recursive graph bisection assigns topically similar documents adjacent ids, shortening the runs a postings scan must skip~\cite{dhulipala2016}. Section~\ref{sec:bmw} returns to this technique.

The closest architecture is \emph{selective search}: partition the corpus into topic-based shards, select a few shards per query, and search only those, exactly~\cite{kulkarni2015}. Hierarchical BM25 shares that skeleton---including its resource bounds, which come with sharding itself, and comparable within-shard scoring, routine in the cooperative single-owner setting. The difference is the selection evidence. CORI scores shards from per-term collection statistics---document counts per term, combined by query operators---and stores no document-level information at all~\cite{callan1995}. Taily models, per shard, the score distribution of the documents containing all query terms, but \emph{estimates} how many such documents exist from single-term counts under an explicit term-independence assumption, adopted precisely to avoid counting mutual term occurrences~\cite{aly2013}. Neither can tell whether the query's terms hit the same document or different documents inside a shard.

ReDDE is the partial exception: it ranks documents in a sampled central index and credits their source shards~\cite{si2003}, so it does observe same-document co-occurrence---but only for documents in the sample, typically 1--4\% of each shard~\cite{aly2013}, and a shard's few documents that co-locate several rare query terms are usually not among them. The evaluations that established these selectors also sat where such limits cost little---mean query lengths of 2.1--3.1 terms for Taily and 3--7 for ReDDE; CORI's 1995 evaluation did use long structured INQUERY queries, but per-term statistics discard co-occurrence at any query length.

At the 16--32-term queries of retrieval-augmented pipelines, same-document co-occurrence separates the right shard from the wrong one, and the $B(c,Q)$ signal of Section~\ref{sec:refinement} tracks it exactly, for every document rather than a sample, for the informative terms whose scattered occurrences raise no single cluster's aggregate score. At query time the signal turns into a selection decision as follows: the query's discriminative terms are looked up in a small inverted index whose postings carry cluster ids; merging those posting lists by document id reveals which documents contain several of the terms at once, and yields, per cluster, the strongest such document's combined idf; that per-cluster score is added to the cluster's aggregate score---the idf-weighted total frequency of the query's terms across all of the cluster's documents---and the top clusters are selected on the sum. Cluster selection thus rests on exactly two measures: how much of the query's vocabulary a cluster holds in aggregate, and whether any single document inside it brings the discriminative terms together. A cluster holding one document where the query's rare terms meet outranks a cluster where the same terms appear only scattered---the distinction every selector above misses. An effectiveness comparison against these selectors remains open, alongside the BMW benchmark (Section~\ref{sec:discussion}).

Tables~\ref{tab:bmwcompare} and~\ref{tab:refinement} summarize the landscape on the axes the paper turns on: whether the top-$k$ is exact, what signal each method needs to save work, and how cost moves as query length $q$ grows. Table~\ref{tab:refinement} separates the design's two selection signals (Section~\ref{sec:refinement}).

\begin{table}[t]
\centering
\caption{Three strategies for a long disjunctive BM25 query. Exhaustive ranked-OR scores the full posting-list union and is exact by construction. BlockMax-WAND (BMW) returns the same top-$k$ but skips documents it can prove cannot enter it; that skipping works only where multi-term co-occurrence exists in the data. Hierarchical BM25 gives up exactness and visits a fixed number of topically selected clusters, so its cost is set by a budget, not by the data. Sections~\ref{sec:wandq} and~\ref{sec:clusterq} develop the last two rows quantitatively.}
\label{tab:bmwcompare}
\footnotesize
\begin{tabular}{@{}lccc@{}}
\toprule
 & \textbf{Ranked-OR} & \textbf{BMW} & \textbf{Hier.\ BM25} \\
\midrule
Exact top-$k$ & yes & yes & no \\
\addlinespace
Work avoided & none & \begin{tabular}{@{}c@{}}provably-losing\\documents\end{tabular} & \begin{tabular}{@{}c@{}}unselected\\clusters\end{tabular} \\
\addlinespace
Signal needed & --- & \begin{tabular}{@{}c@{}}multi-term\\co-occur.\end{tabular} & \begin{tabular}{@{}c@{}}single-term\\concentration\end{tabular} \\
\addlinespace
Cost as $q$ grows & \begin{tabular}{@{}c@{}}grows\\(union)\end{tabular} & \begin{tabular}{@{}c@{}}grows\\(data-dep.)\end{tabular} & \begin{tabular}{@{}c@{}}fixed\\(budget)\end{tabular} \\
\bottomrule
\end{tabular}
\end{table}

\begin{table}[t]
\centering
\caption{The same field re-sorted by co-occurrence evidence: where each method's evidence comes from, and what its query cost tracks as query length grows. ``$T_M$'' is the fixed set of discriminative-but-spread-out terms of Section~\ref{sec:refinement}: high idf, yet scattered across clusters so that no single cluster's aggregate score reflects them; ``agg.'' is selection by the aggregate signal alone (Section~\ref{sec:level1}); ``+$B$'' is the full design, adding the exact same-document co-occurrence signal $B(c,Q)$ over $T_M$. Only the two rightmost designs keep cost independent of query length; the rightmost additionally recovers co-occurrence for $T_M$ terms, which reduces---but does not eliminate---the aggregate signal's dilution, since $A$'s noise from terms outside $T_M$ persists in the combined score.}
\label{tab:refinement}
\scriptsize
\begin{tabular}{@{}lcccc@{}}
\toprule
 & \textbf{Ranked-OR} & \textbf{BMW} & \textbf{Hier.\ (agg.)} & \textbf{Hier.\ (+$B$)} \\
\midrule
\begin{tabular}{@{}l@{}}Co-occurrence\\signal\end{tabular} & \begin{tabular}{@{}c@{}}not needed\\(scores all)\end{tabular} & \begin{tabular}{@{}c@{}}exact,\\all terms\end{tabular} & none & \begin{tabular}{@{}c@{}}exact,\\$T_M$ terms\end{tabular} \\
\addlinespace
\begin{tabular}{@{}l@{}}Query cost\\driver\end{tabular} & \begin{tabular}{@{}c@{}}posting\\union\end{tabular} & \begin{tabular}{@{}c@{}}candidate\\pool\end{tabular} & \begin{tabular}{@{}c@{}}fixed\\budget\end{tabular} & \begin{tabular}{@{}c@{}}budget +\\$T_M$ df\end{tabular} \\
\addlinespace
Degrades as $q \uparrow$ & cost & cost & SNR & \begin{tabular}{@{}c@{}}SNR\\(reduced)\end{tabular} \\
\bottomrule
\end{tabular}
\end{table}

BEIR is the standard zero-shot benchmark for retrieval quality~\cite{thakur2021}; we do not use it. The contribution here concerns latency and memory at $10^9$ documents, beyond BEIR's largest datasets. Section~\ref{sec:eval} measures those two axes directly, and Section~\ref{sec:discussion} confronts the resulting quality gap explicitly.

\section{Hierarchical BM25}
\label{sec:hbm25}

The design target is a hard guarantee: every lexical query returns in under one second over one billion documents, without provisioning hundreds of gigabytes of RAM per node. Hierarchical BM25 meets it by replacing one flat index with two indexes at different granularities. Only a small, bounded fraction is resident at query time.

The design mirrors how a person searches a library. Nobody scans every shelf. You first read the aisle signs and pick the two or three aisles whose labels match your topic. Only then do you search those shelves carefully. Level-1 is the aisle signs: small enough to keep entirely in memory, consulted on every query. Level-2 is the shelf search: large, kept on NVMe, and touched only for the aisles that survived the first step. The trade from Section~\ref{sec:trade} lives entirely in that first step---if the right book sits in an aisle whose sign never mentions your topic, you will not find it. Everything after the first step is exact.

\begin{figure}[t]
\centering
\begin{tikzpicture}[
  font=\scriptsize,
  box/.style={draw, rounded corners=2pt, align=center, inner sep=4pt},
  arr/.style={-{Stealth[length=2mm]}, thick}
]
\node[box, fill=chier!10] (query) {query};
\node[box, fill=chier!20, right=0.55cm of query] (l1) {\textbf{Level-1}\\ ${\sim}$1K groups\\ ${\sim}$4 GB\\ resident};
\node[box, fill=chier!20, right=0.55cm of l1] (l2) {\textbf{Level-2}\\ 1B documents\\ 1M cache\\ + NVMe};
\node[box, fill=chier!10, right=0.55cm of l2] (out) {top-$k$\\ candidates};
\draw[arr] (query) -- (l1);
\draw[arr] (l1) -- node[above]{prune} (l2);
\draw[arr] (l2) -- (out);
\node[box, dashed, below=0.45cm of l1.south west, anchor=north west, text width=6.4cm, fill=cflat!30]
  (flat) {\textbf{Flat index (no hierarchy):} 1B documents, ${\sim}$400 GB, disk-bound};
\node[above=0.12cm of l1.north east, anchor=south, font=\scriptsize\itshape]
  {Hierarchical BM25 (${\sim}$4.4 GB resident)};
\end{tikzpicture}
\caption{Hierarchical BM25. A resident coarse index over document groups prunes the corpus before a fixed-size, NVMe-backed fine index scores survivors, cutting the resident footprint from ${\sim}400$\,GB to ${\sim}4.4$\,GB.}
\label{fig:arch}
\end{figure}
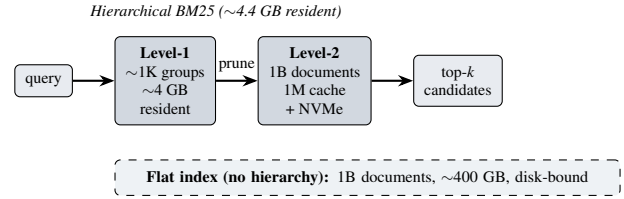

\begin{table}[t]
\centering
\caption{Flat vs.\ hierarchical lexical index. The hierarchy cuts resident memory ${\sim}90\times$, which is what makes the latency guarantee structural rather than average-case. The global DF table (Section~\ref{sec:level1}) lets every cluster score with the same corpus-wide statistics.}
\label{tab:memory}
\footnotesize
\begin{tabular}{@{}llll@{}}
\toprule
\textbf{Configuration} & \textbf{Granularity} & \textbf{Size} & \textbf{Residency} \\
\midrule
Flat (no hier.) & 1B documents & ${\sim}$400 GB & disk-bound \\
Hier.\ Level-1 & ${\sim}$1K groups & ${\sim}$4 GB & resident \\
Hier.\ Level-2 & 1B documents & \begin{tabular}{@{}l@{}}${\sim}$400 GB total\\(${\sim}$400 MB res.)\end{tabular} & 1M cache + NVMe \\
Global DF table & $|V|$ terms & ${\sim}$100 KB & resident \\
\bottomrule
\end{tabular}
\end{table}

\subsection{Why a flat index fails at a billion documents}

A flat inverted index stores, per term, the document frequency and per-document keyword frequencies BM25 needs. At one billion documents this is ${\sim}400$\,GB. Holding it resident is possible on large-memory servers, but at a DRAM cost that is both large and linear in the corpus (Section~\ref{sec:positioning}); the economical deployment---and the baseline we measure---serves it from disk. The postings for a multi-term query are scattered, so each query issues many random reads. Latency then scales with fan-out rather than with the number of relevant documents. Tail latency routinely exceeds one second. This is the ``no-hierarchy'' baseline.

\subsection{A two-level index}

Hierarchical BM25 organizes the corpus into two levels (Figure~\ref{fig:arch}, Table~\ref{tab:memory}).

\textbf{Level-1 (document groups).} Documents are aggregated into ${\sim}1$K groups. The Level-1 index carries group-level document- and keyword-frequency statistics. It is ${\sim}4$\,GB, pre-computed, and resident. It acts as a coarse filter: it identifies the groups most likely to contain relevant documents, before any fine scoring runs.

\textbf{Level-2 (documents).} The Level-2 index holds fine per-document statistics for all one billion documents. We do not pin it in memory. Instead, we serve it through a fixed-size cache of ${\sim}1$M entries (${\sim}400$\,MB resident). The remainder lives on NVMe and is paged in on demand. The cache size is fixed and independent of corpus size. The resident footprint---and therefore the worst-case query cost---does not grow as the corpus grows.

Section~\ref{sec:clustering} describes how the ${\sim}1$K groups are formed in the first place; Section~\ref{sec:level1} then details how we build Level-1 over them and how we make scores from different Level-2 clusters comparable before merging.

\subsection{Forming the clusters: balanced topical LDA}
\label{sec:clustering}

How the corpus is partitioned into ${\sim}1$K groups is not a free choice. It is the assumption everything downstream stands on. Level-1 is selective only if term statistics differ sharply between clusters: if documents were grouped by ingestion order or by hash, every cluster's vocabulary would approximate the global distribution, the ${\sim}1$K per-cluster term statistics of Section~\ref{sec:level1} would be nearly interchangeable, and selecting the top 40 would be close to arbitrary---precisely where the recall surrendered by the trade would collapse. In the library analogy: aisle signs help only if books are shelved by subject. A library shelved by arrival date has signs, but every sign reads the same. We therefore partition \emph{topically}, using Latent Dirichlet Allocation (LDA)~\cite{blei2003} with two deliberate constraints: which terms the model sees, and how large a cluster is allowed to grow.

\textbf{Features: the mid-frequency band only.} Luhn's classic observation, built on Zipf's law, is that a term's power to discriminate content peaks at mid frequency~\cite{luhn1958}. The highest-frequency terms appear everywhere and say nothing about topic---knowing a document contains ``the'' or ``system'' places it on no particular shelf. The rarest terms are individually informative but statistically useless for clustering: a term appearing in a few dozen documents out of a billion gives a topic model almost nothing to generalize from, and produces sparse, unstable assignments. So each document's feature vector is restricted to the ${\sim}$10K mid-idf terms of the $|V| \approx 20{,}680$-term vocabulary, dropping the highest-df head and the rare tail. This band overlaps the discriminative-but-spread-out terms that the co-occurrence signal tracks (Section~\ref{sec:refinement}); both mechanisms want terms discriminative enough to carry topical signal, but the two select differently within it: clustering features favor terms that \emph{concentrate} (they define where a document lives), while the co-occurrence signal's $T_M$ deliberately favors the discriminative terms that \emph{do not} concentrate, since those are exactly the ones no single cluster's score can surface. The vocabulary's topical middle feeds both, split by whether a term localizes or spreads.

\textbf{Balance is enforced, not hoped for.} The latency bound assumes the top-$k_{\mathrm{clu}} = 40$ clusters cover ${\sim}4\%$ of the corpus. Topic popularity is itself Zipfian, so assigning each document to its single most probable topic would produce a few enormous clusters---one popular-topic cluster holding 10--20\% of all documents, which breaks the budget the moment a query selects it. No aisle may hold a fifth of the library. We therefore use LDA with a modest topic count for representation (each document gets a topic-proportion vector) and impose balance at assignment: documents are grouped into ${\sim}1$K equal-size clusters over those topic vectors, via capacity-capped assignment with overflow splitting. A document capped out of its best-fitting cluster goes to its next-best; the cost is a small loss of topical purity at cluster boundaries, paid to keep every cluster's size---and therefore per-query work---bounded. This is the build-time twin of the query-time guarantee in Section~\ref{sec:queryproc}: the fixed cluster budget bounds cost only because no single cluster can be oversized.

\textbf{Growing $K$ by splitting, without reclustering.} The cluster count is not frozen at build time. As the corpus grows or its topic mix drifts, individual clusters cross the size threshold that the balance invariant protects: a cluster exceeding, say, $1.5\times$ the target size $N/K$ inflates the per-query work of any query that selects it. Rather than re-running LDA over the whole corpus to raise $K$ globally, we split only the offending clusters, locally. Because every document already carries its LDA topic-proportion vector from the representation stage, a split needs no new clustering pass: we partition the oversized cluster's documents by their dominant \emph{secondary} topic (the strongest topic other than the one that placed them here), sending each sub-topic to a child cluster. This is a partition of vectors already in hand, $O(|\text{cluster}|)$ and local to one cluster's documents. Three properties make it cheap in the parts that matter. The two child Level-2 indexes are re-partitioned from the parent's postings alone, touching no other cluster. The two children's Level-1 statistics are recomputed from each child's own vocabulary (Section~\ref{sec:level1})---two small rows in the resident table. And the global DF table is \emph{untouched}: $\mathrm{df}(t) = \sum_c \mathrm{df}_c(t)$ splits the parent's contribution across two children without changing the total, so the cross-cluster scoring correction of Section~\ref{sec:level1} needs no update and Level-2 scoring stays exactly correct through the split. Splits are per-cluster and independent, so they parallelize and can run online while the rest of the index serves. The one thing splitting does \emph{not} do is restore global optimality: a document trapped in a lineage stays there, since splits subdivide but never migrate documents across the cluster tree. Splitting is the fast incremental path that keeps clusters near target size between the periodic full reclusterings discussed in Section~\ref{sec:discussion}; it defers those rebuilds rather than eliminating them.

One caveat on what this buys. Topical partitioning makes the single-term concentration signal of Section~\ref{sec:bmw} \emph{exist}; it does not by itself say how much recall the top-40 selection preserves. That number---and the gap between this partitioning and cheaper alternatives such as cutting a recursive-graph-bisection ordering~\cite{dhulipala2016} into equal segments---belongs to the Recall@$k$ measurement flagged in Section~\ref{sec:discussion} as the most important open step.

\subsection{Building Level-1 and normalizing across clusters}
\label{sec:level1}

The first of the two selection signals is the aggregate one: the idf-weighted total frequency of the query's terms across a cluster's documents. Level-1 is not a hand-designed summary; it is built mechanically from each cluster's own index, and two details decide whether it works. The first is the per-cluster statistic each term contributes: get it wrong and the coarse filter stops being selective. The second is how independently computed per-cluster scores are made comparable at merge time: get it wrong and the merged ranking stops being valid.

\textbf{The per-cluster term statistic.} For each Level-2 cluster, we scan that cluster's inverted index in parallel across its segments, sum each term's raw occurrence count---its collection frequency $f_c(t)$ within the cluster---and keep the highest-frequency terms. Each kept term contributes one entry to a resident table: a saturated weight

\[
w_c(t) = \left(\log_2 \max(f_c(t),1)\right)^2 .
\]

The saturation matters, in the same spirit as BM25's own term-frequency saturation: a term ten times more frequent contributes roughly four times the weight, not ten, so one very common term cannot dominate a cluster's representation and crowd out the rest of its vocabulary. The whole of Level-1 is this table---per term, a sparse vector of cluster weights---plus each cluster's document count. Nothing else is needed at the coarse level.

One property of this construction matters later. It sums a term's evidence across \emph{every} document in the cluster, rather than taking the best single document. A query term's Level-1 signal in a cluster therefore reflects that term's aggregate presence there. Section~\ref{sec:bmw} returns to this observation when comparing against WAND-style pruning.

\textbf{Making per-cluster scores comparable.} The second specification is subtler. Level-2 is not one shared index but ${\sim}1$K independent per-cluster BM25 indexes. Left to itself, each cluster's search engine infers BM25's inverse-document-frequency term from what that cluster alone contains: its own document count $N_c$ and its own per-term document frequency $\mathrm{df}_c(t)$. This is a correctness bug, not just an approximation. BM25 is additive across terms---$S(d,Q) = \sum_t \mathrm{idf}(t)\,\mathrm{TF}(t,d)$---and a term's true, corpus-wide $\mathrm{idf}(t)$ can differ sharply from its cluster-local estimate, in either direction. Two documents with identical term frequencies can then receive different scores purely because of which cluster they sit in.

The bug is easiest to see as grading on a curve. Two students hand in identical exams. One sits in a class of experts, the other in a class of beginners. Curved, per-class grades score the same exam differently. An absolute standard scores it the same everywhere. Per-cluster IDF is the curve; the global table we introduce below is the absolute standard.

A concrete retrieval example. A document group organized by topic concentrates that topic's characteristic terms. Inside a cluster of Kansas-related documents, a term like ``kansas'' looks common ($\mathrm{df}_c/N_c$ is high) even though it is globally rare. ``Population'' looks unremarkable in both places. Two documents with the same term frequencies for \{kansas, population\}, sitting in different clusters, can end up $1.6\times$ apart in score. No single per-cluster correction factor can repair this after the fact. The distortion is per-term and hidden inside a sum; a single scalar cannot undo per-term errors that do not move together.

We avoid the problem instead of correcting for it. At build time, alongside the Level-1 vocabulary extraction, we aggregate $\mathrm{df}(t) = \sum_c \mathrm{df}_c(t)$ and the global average document length $\mathrm{avgdl}$ across all ${\sim}1$K clusters into one table, keyed by term. The table is small: one count per term, $|V| \approx 20{,}680$ in our setting, ${\sim}100$\,KB total (Table~\ref{tab:memory}). It is the only piece of global state Level-2 scoring needs. At query time, we construct each cluster's BM25 weight from the global $N$, $\mathrm{df}(t)$, and $\mathrm{avgdl}$ in this table---not from the cluster's own inferred statistics. Every other input to the score---raw term frequency $f_{t,d}$, document length $|d|$---is already a correct, local property of the document. Supplying only these three global quantities makes each cluster's score identical to the score a single flat index over the whole corpus would have returned.

Merging the per-cluster top-$k$ lists into one global ranking is then a plain concatenate-and-sort. No per-cluster correction factor, and nothing left to get wrong. The only cost is building and keeping resident a ${\sim}100$\,KB table. It changes neither the postings read per query nor the clusters visited, so it leaves the latency and throughput results in Section~\ref{sec:eval} unaffected. It only changes which number each visited cluster reports as a document's score---and makes that number correct. This fix is also what confines the trade of Section~\ref{sec:trade} to cluster selection alone. Rank safety is surrendered only in deciding which clusters to visit---never in how a visited document is scored.

\subsection{The co-occurrence selection signal}
\label{sec:refinement}

The aggregate score is deliberately only half of cluster selection. The design pairs it with a second signal, built by the same scan, that supplies the one kind of evidence aggregation cannot: whether several of the query's discriminative terms meet in a single document.

The set of terms worth tracking for co-occurrence is not simply ``the important terms.'' It is, more precisely, the terms the aggregate signal $A(c,Q)$ fails to \emph{route}---fails, that is, to place their clusters among the selected ones---because $B$ is redundant wherever $A$ already succeeds. Two properties define that set, and both are necessary.

First, the term must be discriminative: high idf. A term with $\mathrm{idf}(t) \approx 0$ contributes almost nothing to $S(d,Q) = \sum_t \mathrm{idf}(t)\,\mathrm{TF}(t,d)$ wherever it occurs, so two such terms co-occurring in a document means nothing---their combined score is negligible. Tracking co-occurrence among low-idf terms detects coincidences, not matches. Only for high-idf terms does same-document co-occurrence imply a genuine top-$k$ candidate.

Second, and this is the property the aggregate signal makes essential, the term must be \emph{spread across many clusters}, not concentrated in a few. Here the two signals divide the vocabulary cleanly. A \emph{concentrated} discriminative term, one confined to a handful of clusters, already routes correctly under $A$ alone: those few clusters score high on single-term concentration and clear the top-$k_{\mathrm{clu}}$ cutoff without any co-occurrence information. For such a term $B$ adds nothing: its clusters were going to be visited regardless. A \emph{spread-out} discriminative term is the opposite case: scattered thinly across many clusters, it pushes no single cluster over the cutoff, so to $A$ it is nearly invisible, almost a stopword. Its only path to mattering is to co-occur, in one document, with another such term---and detecting that requires exactly the document-level tracking $A$ lacks. These spread-out, high-idf terms are the terms $B$ exists for.

This inverts a natural but wrong intuition. One might select $T_M$ by concentration, reasoning that the most topical terms deserve tracking. But a concentrated term is precisely the term $A$ handles best; tracking its co-occurrence is wasted effort. The terms that need $B$ are the discriminative terms that \emph{refuse} to concentrate---individually meaningful, yet localized nowhere, so that a real match built from two of them is invisible to any aggregate. We therefore define $T_M$ as the $M$ terms scoring highest on the product of discriminativeness and cluster spread: $\mathrm{idf}(t) \cdot H_{\mathrm{clu}}(t)$, where $H_{\mathrm{clu}}(t)$ is the entropy of $t$'s document-frequency distribution across the ${\sim}1$K clusters, high when $t$ is smeared evenly and low when $t$ concentrates. Both factors are already materialized: $\mathrm{idf}(t)$ from the global DF table (Section~\ref{sec:level1}), and the per-cluster $\mathrm{df}_c(t)$ that $H_{\mathrm{clu}}$ needs from the same Level-1 aggregation. The signal adds one cheap question per cluster---does any \emph{single} document here contain several of the query's spread-out discriminative terms together?---which is exactly the question $A(c,Q)$ cannot answer and, for these terms, the only question that routes them.

We therefore build a second, small inverted index over only $T_M$ (we use $M = 1{,}000$ of the $|V| \approx 20{,}680$ terms). Concretely, it is an ordinary inverted index with one addition: each posting stores the pair $(\text{document id}, \text{cluster id})$ rather than the document id alone, and each term's postings are kept sorted by document id. The cluster id riding along in every posting is what lets the query-time computation below attribute a document to its cluster without a second document-to-cluster lookup. The whole index's size is $\sum_{t \in T_M} \mathrm{df}(t)$; because $T_M$ excludes the high-df head---those terms have low idf and are filtered out by the discriminativeness factor---this sum stays bounded, though the spread criterion means the terms are not the very rarest in the vocabulary and their lists are correspondingly longer than a rare-tail selection would give (Section~\ref{sec:clusterq} quantifies the resulting cost). It is built from the same postings scan already used for the rest of Level-1 (Section~\ref{sec:level1}): the cluster id is known at build time, so tagging costs nothing extra and adds no new traversal.

At query time, for the query terms that fall in $T_M$ (call this subset $Q_M = Q \cap T_M$), we retrieve their posting lists, group matches by document, and compute, per cluster $c$, the best co-occurring document's $T_M$-term score:
\[
B(c,Q) = \max_{d \in c} \sum_{t \in Q_M \cap d} \mathrm{idf}(t).
\]
$B(c,Q)$ is exact, not estimated. It directly answers whether this cluster contains one document that hits several of the query's spread-out discriminative terms at once---precisely the question $A(c,Q)$ cannot answer, and, for terms too scattered to raise any cluster's aggregate score, the only question that routes them. A cluster containing a document that matches five terms from $Q_M$ scores higher under $B$ than one whose best document matches two, in proportion to those terms' idf. This matches the intuition: co-occurrence among discriminative terms should move the ranking, while co-occurrence among terms like ``consists'' or ``together'' should not---and indeed cannot, since such terms carry negligible idf and are excluded from $T_M$ in the first place.

\textbf{Selecting clusters at query time.} Selection computes both signals over resident structures and ranks their sum. For the aggregate signal, we tokenize the query and, for each query term, fetch its row of cluster weights from the routing table; every cluster accumulates $A(c,Q) = \sum_{t \in Q} \mathrm{idf}(t)\,w_c(t)$---$q \times {\sim}1$K multiply-adds, negligible next to everything downstream. A cluster holding evidence for several query terms therefore outranks one that matches a single term very strongly: the aggregate half rewards coverage of the query, not a single lucky match. For the co-occurrence signal, the query's $T_M$ terms drive the posting-list merge described next, yielding $B(c,Q)$ per cluster. The two combine additively:
\[
\mathrm{Score}(c,Q) = A(c,Q) + \lambda\, B(c,Q).
\]
Retaining $A(c,Q)$ means a cluster still earns credit for single-term concentration among the $q - |Q_M|$ terms outside $T_M$; $B(c,Q)$ contributes the document-level evidence for the terms most likely to determine the true top-$k$. $\lambda$ is a tunable weight; both terms are already idf-scaled, so $\lambda = 1$ is a reasonable starting point pending empirical tuning. Only the top-scoring clusters proceed to Level-2, not the whole ${\sim}1$K. Section~\ref{sec:eval} uses 40.

\textbf{Computing $B$ at query time.} The definition of $B(c,Q)$ takes a max, over documents in a cluster, of a per-document idf sum---so the computation must group $T_M$-term hits \emph{by document} first, then reduce to a per-cluster max. Because the $T_M$ index keeps each posting list sorted by document id, this is a document-at-a-time merge, the same traversal an ordinary inverted index uses for a conjunctive query. We advance a cursor into each of the $|Q_M|$ query-$T_M$ lists, always stepping the one at the smallest document id; galloping search over the sorted lists keeps the merge efficient when the lists differ sharply in length. Two small hash maps hold the running state: one keyed by document id accumulates that document's idf sum as each list contributes a hit, and one keyed by cluster id keeps the largest per-document sum seen so far in that cluster. When a document is finalized (all cursors have passed it), its accumulated sum updates its cluster's running max---and the cluster id needed for that update is already in the posting, so no separate lookup is required. After the merge, the cluster map holds $B(c,Q)$ for every cluster the $T_M$ query terms touched; all other clusters take $B = 0$. Both maps are bounded by the number of distinct documents the $Q_M$ terms reach, i.e.\ $\sum_{t \in Q_M} \mathrm{df}(t)$---the same quantity that bounds query cost below.

Two structural choices keep this cheap rather than merely correct. First, we deliberately do \emph{not} precompute a cluster-by-term co-occurrence table: such a table would store each cluster's \emph{aggregate} $T_M$-term presence, which is exactly the document-blind signal $A(c,Q)$ already carries and $B$ exists to go beyond; it cannot tell whether two such terms landed in the same document or two different ones. Second, we do not materialize pairwise term-term co-occurrence counts either; those are $O(|T_M|^2)$, mostly empty, and answer a corpus-wide question rather than the per-cluster, per-document one $B$ needs. The plain cluster-tagged inverted index avoids both extremes: it stores document-level evidence, so the merge can see same-document co-occurrence, without precomputing any pair, so storage stays linear in $\sum_{t \in T_M} \mathrm{df}(t)$. If the per-cluster max reduction ever dominates, a secondary posting sort (cluster id major, document id minor) lets the merge finish one cluster at a time and bound the document-keyed map to a single cluster's documents, trading a little build-time sort for a smaller query-time working set.

\subsection{Query processing and the guarantee}
\label{sec:queryproc}

A query is answered in two passes. First, the resident selection structures score the ${\sim}1$K groups---the aggregate signal from the routing table plus, for the query's $T_M$ terms, the co-occurrence signal of Section~\ref{sec:refinement}---and prune to the promising ones. This pass is cheap: the structure is small and in memory. Second, only the survivors are scored against Level-2, whose postings come from the fixed-size cache or are paged in from NVMe.

\textbf{Serving the per-cluster indexes.} Rather than reopening a cluster's Level-2 index from disk on every query, we open a handle---index, reader, and schema---for every cluster once, ahead of query serving, and reuse those handles across queries. The underlying index files are memory-mapped. Postings for frequently accessed clusters therefore stay resident under ordinary OS page-cache pressure, while postings for cold clusters remain on NVMe until a query actually reaches them. This is the mechanism that realizes the fixed memory budget in Table~\ref{tab:memory} in aggregate, without an explicit least-recently-used table keyed by individual postings.

\begin{figure*}[t]
\centering
\begin{tikzpicture}
\node[plain, text width=2.0cm] (q) at (0,0) {Query\\ (16--32 terms)};
\node[op, text width=4.2cm] (A) at (5.1, 1.45) {signal $A$: which clusters\\ are rich in the query's terms?};
\node[op, text width=4.2cm] (B) at (5.1,-1.45) {signal $B$: where do rare terms\\ meet in a single document?};
\node[res, text width=2.2cm] (sel) at (9.9,0) {${\sim}$40 clusters\\ selected};
\node[disk, text width=2.3cm] (ex) at (12.85,0) {exact BM25\\ inside each};
\node[plain, text width=1.5cm] (topk) at (15.55,0) {global\\ top-$k$};
\draw[arr] (q.east) -- ++(0.55,0) |- (A.west);
\draw[arr] (q.east) -- ++(0.55,0) |- (B.west);
\draw[arr] (A.east) -| node[dlab, pos=0.92, anchor=west, xshift=1.5mm] {$+$} (sel.north);
\draw[arr] (B.east) -| (sel.south);
\draw[arr] (sel) -- (ex);
\draw[arr] (ex) -- (topk);
\end{tikzpicture}
\caption{\textbf{Query time, conceptually.} Routing first, exactness after. Two complementary signals pick the clusters: $A$ asks which clusters are rich in the query's terms overall, and $B$ asks where several discriminative terms meet in one document---the evidence $A$ cannot see. Only the ${\sim}$40 selected clusters are searched, with exact BM25 scores throughout, so the approximation lives entirely in the selection step. (Mechanics in Sections~\ref{sec:queryproc} and~\ref{sec:refinement}.)}
\label{fig:phase3}
\end{figure*}
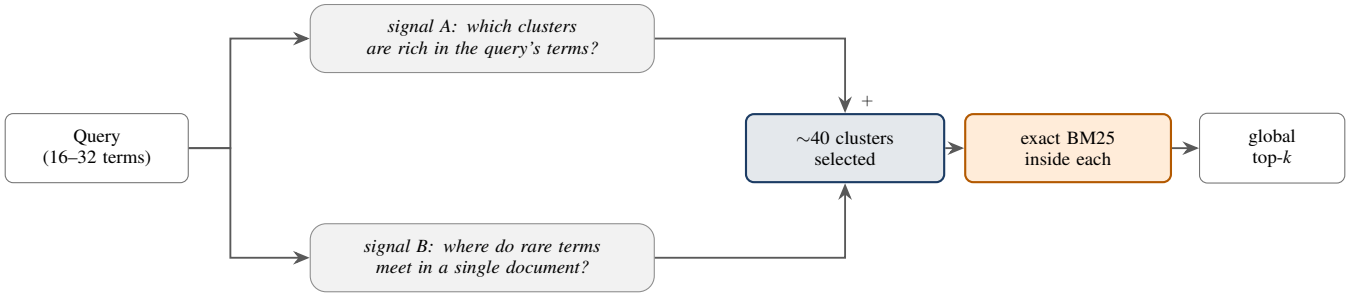

Two facts make the sub-second bound structural rather than merely average-case at this operating point. The Level-1 pass bounds how many documents reach fine scoring. The fixed-size Level-2 cache bounds resident memory and worst-case paging. The two bounds have different scopes, and it matters. The \emph{memory} bound is corpus-independent: the resident footprint stays ${\sim}4.4$\,GB whether the corpus is one billion documents or larger, because nothing resident grows with $N$. The \emph{latency} bound is not: at a fixed cluster count, a growing corpus puts more documents inside each visited cluster, so per-query work grows with $N/K$ (Section~\ref{sec:discussion}); holding it requires growing $K$, incrementally via the splits of Section~\ref{sec:clustering}. The sub-second guarantee is therefore a property of the one-billion-document, ${\sim}1$K-cluster configuration measured in Section~\ref{sec:eval}, maintained under growth by rebalancing---not an asymptotic invariant.

The particular split---${\sim}1$K groups at Level-1, a 1M-entry cache at Level-2---is not arbitrary. Fewer, larger groups would shrink Level-1 further but prune less precisely, pushing more documents into the expensive disk-backed pass. More, smaller groups approach the flat baseline and re-introduce its memory cost. ${\sim}1$K groups is the point at which the coarse index is cheap enough to keep fully resident, yet selective enough that only a small fraction of the corpus survives to Level-2. The Level-2 cache size follows the same logic in the other direction. It is sized to the working set of frequently accessed postings under realistic query skew, not to the corpus. Growing the corpus grows only the cold tail on NVMe, not the resident footprint.

\section{Comparison to BlockMax-WAND at Long Query Lengths}
\label{sec:bmw}

Retrieval-augmented pipelines issue longer queries than the 2--5-term web queries WAND-family pruning was originally evaluated against---often 16--32 terms, once a query is combined with expansion terms, related fields, or metadata filters expressed as additional disjuncts. This section argues that this regime specifically favors cluster-level selection over WAND-style pruning. It gives a quantified mechanism for why, and states plainly what remains unmeasured.

\subsection{Why WAND's pruning power depends on query length}
\label{sec:wandq}

Section 2 introduced the WAND-family mechanism and its high-jump intuition: a running score threshold $\theta$---the current $k$-th best score---lets the pivot advance past any run of documents whose upper-bound score cannot exceed it. What that summary leaves open is quantitative. The size of each skip is governed by how far apart, in document-id order, the matching documents are. Sparse candidates permit large skips. Dense candidates force the pivot to advance one small step at a time.

For a term with document frequency $\mathrm{df}(t)$ over $N$ documents, the probability that an arbitrary document matches at least one of $q$ independent query terms of similar frequency $p = \mathrm{df}(t)/N$ is
\[
P(\text{matches} \geq 1) = 1 - (1 - p)^q.
\]
Using this corpus's own term statistics ($N = 10^9$, $|V| \approx 20{,}680$, ${\sim}400$\,GB of keyword-frequency data $\Rightarrow$ average $\mathrm{df}(t) \approx 5 \times 10^6$, $p \approx 0.005$), the candidate pool---the set of documents WAND's pivot must at least consider---grows substantially with query length:

\begin{center}
\footnotesize
\begin{tabular}{@{}lccc@{}}
\toprule
$q$ (terms) & 8 & 16 & 32 \\
\midrule
Candidates matching $\geq 1$ term & 39M & 77M & 148M \\
\bottomrule
\end{tabular}
\end{center}

Longer queries do raise $\theta$ faster in principle. More terms give more ways for a document to accumulate score. But this holds only when query terms \emph{correlate}---when the same documents that match one term are disproportionately likely to match others. For topically unrelated terms drawn independently, the true top-$k$ requires rare multi-term co-occurrences scattered across the corpus. $\theta$ then climbs slowly while the candidate pool grows as shown above. This is the opposite of WAND's favorable regime (short queries, correlated terms)---and it is the mechanism behind MaxScore's robustness on long disjunctive queries noted in Section 2, since MaxScore does not pivot through the dense docid stream at all. A fair baseline for this comparison should therefore include MaxScore; we have not yet benchmarked it.

\subsection{Why cluster selection does not have the same growth driver}
\label{sec:clusterq}

An everyday version of the contrast first. Finding one person who speaks all 16 languages on a list is hard: such people are rare, and you must check candidates one by one. Finding a neighborhood where each of the 16 languages is spoken by \emph{someone} is easy: many districts qualify. WAND's pruning accelerates only when the first kind of match exists. Cluster selection needs only the second.

In index terms: Level-1 selection (Section~\ref{sec:level1}) needs a structurally weaker signal than WAND's pivot condition. WAND's threshold rises only when a single document accumulates score from several query terms at once, that is, document-level co-occurrence. Level-1's per-cluster score for a query term, by contrast, sums that term's evidence across every document in the cluster (Section~\ref{sec:level1}). It requires only that the term be well represented \emph{somewhere} in the cluster. No other query term needs to co-occur with it in the same document. A cluster can be correctly identified as promising because it is rich in one of the 16--32 query terms, even if the other terms are topically unrelated to it. In exactly the regime where WAND struggles, this signal is more available. The signal does not occur naturally: the balanced topical clustering of Section~\ref{sec:clustering} creates it at build time. Under an arbitrary partition, no cluster would be richer in any term than any other, and this section's argument would have nothing to stand on.

Architecturally, Level-1's cost also does not grow with query length the way WAND's candidate pool does. The number of clusters visited is fixed by the top-$k_{\mathrm{clu}}$ selection policy (Section~\ref{sec:level1}), not by how many documents happen to match a term. Table~\ref{tab:bmwcompare} (Section~\ref{sec:positioning}) states this three-way contrast in full.

This picture is incomplete in one important way. The aggregate signal that gives Level-1 its query-length independence has a cost of its own. Recall the cluster score $A(c,Q) = \sum_{t \in Q} \mathrm{idf}(t)\,w_c(t)$, where $w_c(t)$ is term $t$'s saturated aggregate weight in cluster $c$ (Section~\ref{sec:level1}). As $q$ grows, this sum is increasingly dominated by terms with no special relationship to any one cluster. Suppose only a handful of query terms are strongly cluster-discriminative. Their contribution to $A(c,Q)$ stays roughly fixed as $q$ grows, while the noise from the remaining terms grows with $q$; under a rough independence assumption, its variance scales linearly in the number of such terms. $A(c,Q)$'s signal-to-noise ratio therefore falls at roughly $1/\sqrt{q}$. At $q = 32$, this can wash out the very topical concentration this section's argument relies on. The intuition: judging a cluster by summing 32 terms' evidence, when only three terms are actually informative about it, is like rating a restaurant by averaging 32 reviews when only three reviewers ate there. The informative few are diluted by the uninformative many.

A second, sharper problem: $A(c,Q)$ cannot tell whether several query terms' evidence came from the \emph{same} document or from several different documents scattered across the cluster. That distinction is exactly what determines whether the cluster actually contains a genuinely strong match. The co-occurrence signal (Section~\ref{sec:refinement}) exists for precisely these two failure modes; Section~\ref{sec:bcost} prices it against BMW.

\subsection{The co-occurrence signal against BMW's candidate pool}
\label{sec:bcost}

The co-occurrence signal's cost profile is what makes it viable exactly where BMW degrades. Query cost is bounded by $\sum_{t \in Q_M} \mathrm{df}(t)$---the total postings touched across the query's $T_M$ terms. These terms are not the rarest in the vocabulary, so their lists are longer than a rare-tail selection would give; the cost is therefore higher than tracking the rarest $M$ terms, but still bounded, because the discriminativeness factor excludes the high-df head entirely. Critically, this cost does not grow with $q$ the way BlockMax-WAND's candidate pool does. It grows with how many of the query's terms fall in $T_M$, not with $q$ itself. A 32-term query with no $T_M$ terms costs the same as an 8-term query with none. WAND's candidate pool, by contrast, grows from 39M to 148M documents across that same range, regardless of which specific terms are involved.

Positioned this way, the two-signal selection is better suited than either pure alternative at the query lengths this paper is concerned with. Relative to the aggregate signal alone, the co-occurrence signal supplies exactly the two kinds of evidence Section~\ref{sec:clusterq} showed missing, at a cost bounded by how many $T_M$ terms a query contains rather than by $q$. Against BlockMax-WAND, it recovers real, exact, document-level co-occurrence---the same signal BMW's own pivot condition relies on---restricted to the subset of terms where tracking it is cheap. Its query cost is governed by how many $T_M$ terms a query happens to contain, not by the corpus-wide candidate pool a disjunctive query touches. It should therefore avoid BMW's degradation on long, weakly-correlated queries while adding back the one signal the pure aggregate score could not detect---though it mitigates rather than eliminates the dilution of Section~\ref{sec:clusterq}, since $A$'s noise from the $q - |Q_M|$ terms outside $T_M$ remains in the combined score. Table~\ref{tab:refinement} (Section~\ref{sec:positioning}) places the two-signal selection alongside exhaustive ranked-OR, BMW, and the aggregate-only baseline on exactly these dimensions.

Three further limits apply to the co-occurrence signal specifically, on top of the two already stated for cluster selection in general. First, it helps only when a query actually contains terms from $T_M$, \emph{and} when at least two of them co-occur in one document within a cluster that $A$ would otherwise rank below the cutoff. This is a genuinely narrow window, since a single $T_M$ term contributes nothing new (with one term there is no co-occurrence to detect). On this paper's own evaluation queries (terms drawn uniformly from the full 20,680-word dictionary), the expected number of $T_M$ terms in a query is small and the two-term co-occurrence event smaller still, so the signal fires rarely on the benchmark in Section~\ref{sec:eval}. It helps instead on realistic or expanded queries that concentrate on discriminative terms---plausibly including a companion query-expansion method's own output, since attribution-selected expansion terms are themselves chosen for high discriminative value, and are exactly the spread-out discriminative terms $T_M$ targets. Second, $M$ and the idf-versus-spread weighting in the $T_M$ score both trade coverage against storage and query cost, and we have not swept them. Third, and most fundamentally, $\mathrm{Score}(c,Q)$ is still not a bound of any kind. Nothing prevents a cluster with a low combined score from containing a genuine top-$k$ document whose relevant terms all fall outside $T_M$. This remains, like the rest of Section~\ref{sec:hbm25}, an approximate heuristic rather than a rank-safe one. Its contribution to selection quality has not been isolated empirically; that measurement joins the open list of Section~\ref{sec:discussion}.

\subsection{What this argument does and does not establish}

Two limits on this argument matter as much as the mechanism itself.

First, this is not an equal-accuracy comparison; it is the price side of the trade from Section~\ref{sec:trade}, stated plainly. BMW is rank-safe: it returns exactly the top-$k$ an exhaustive scan would. Hierarchical BM25's cluster selection carries no such guarantee. A cluster with a low selection score can still contain a document that would rank in the global top-$k$. Section~\ref{sec:quality} measures how much this costs at a 500K-document configuration---visiting 5--10\% of clusters recovers 0.83--0.92 of the exhaustive result score---though not yet at billion scale. The comparison in this section therefore remains about cost given approximate operation, not cost at equal quality. Section~\ref{sec:discussion} returns to this limitation.

Second, document reordering narrows the gap this section relies on. WAND's density problem is a property of document-id \emph{order}, not an inherent limit: the recursive graph bisection reordering described in Section 2 makes a term's postings locally dense in id-space, restoring much of WAND's ability to make large skips. Cluster selection and document reordering are, at bottom, two different data structures for exploiting the same topical signal. A fair resolution of this comparison therefore requires benchmarking against BlockMax-WAND \emph{with} such reordering---something we have not done. Until then, this section's conclusion should be read as a mechanism-level hypothesis with analytical support, not as a settled result.

\section{Evaluation}
\label{sec:eval}

\subsection{Setup}
\label{sec:setup}

We measure Hierarchical BM25 in isolation: no query expansion, no semantic stage. The corpus is one billion documents, partitioned into ${\sim}1$K balanced topical clusters (Section~\ref{sec:clustering}). The hardware is a single node with 64 AMD EPYC 7343 CPU cores and eight Intel P5500 NVMe SSDs in a RAID-0 array. The index is read from disk rather than pre-warmed into the page cache, so these numbers reflect cold-start I/O, not a best case.

Single-query latency is measured with 8-, 16-, and 32-term queries built from terms drawn at random from the corpus dictionary. This mix is deliberately adversarial. It is designed to maximize postings fan-out, not to resemble a natural query, because fan-out is what stresses the index structure. At this setting, Hierarchical BM25 visits about 4\% of the corpus per query---the fraction of documents whose Level-1 clusters survive the coarse filter.

One property of this corpus must be stated for the term-statistics arguments to be interpreted correctly. Its vocabulary is compact ($|V| \approx 20{,}680$ terms over $10^9$ documents), so every term is frequent (average $\mathrm{df}(t) \approx 5 \times 10^6$) and there is no natural rare tail. This is well suited to what the benchmark measures: uniform-random queries over a compact, uniformly frequent vocabulary maximize postings fan-out, which is the structural stress the latency and throughput claims are about. It is poorly suited to everything else: a natural corpus at this scale carries a vocabulary orders of magnitude larger with a Zipfian tail, and the clustering-feature selection of Section~\ref{sec:clustering} and the $T_M$ selection of Section~\ref{sec:refinement} are specified for that setting and exercised only weakly here. The same uniform-random query mix is also maximally adversarial to the paper's own quality mechanisms, since it neutralizes topical routing and $T_M$ alike; nothing about retrieval quality should be inferred from this setup, in either direction. Validation on a natural-vocabulary corpus is open.

We compare against two flat (non-hierarchical) baselines built over the same one billion documents: a single-threaded flat index, and a multi-threaded flat index (Flat-MT) that parallelizes the same disk-bound scan across cores. Both implement exhaustive ranked-OR (Section 2): every document matching at least one query term is scored, so their per-query cost tracks the posting-list union sizes tabulated in Section~\ref{sec:wandq}. We do not compare against BlockMax-WAND empirically in this evaluation. Section~\ref{sec:bmw}'s comparison is analytical, and closing that gap is the most direct next step for this work.

\begin{figure}[t]
\centering
\includegraphics[width=0.86\columnwidth]{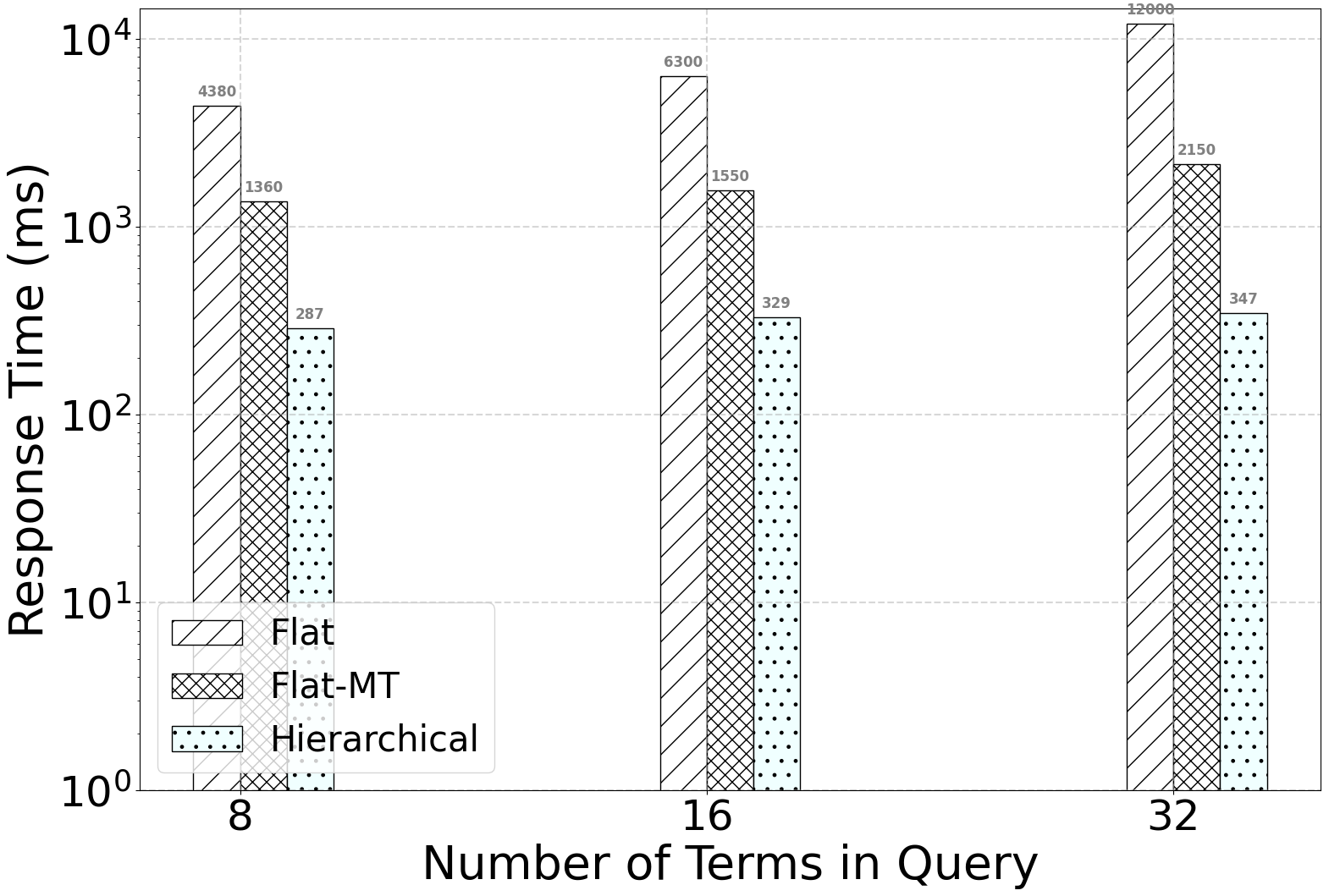}
\caption{Measured single-query latency (log scale) over 1B documents in 1K clusters, with ${\sim}4\%$ of the corpus visited per hierarchical query; 8/16/32-term queries. Hierarchical BM25 stays near 300\,ms while the flat and flat multi-threaded indexes run 4--30$\times$ slower and breach the 1\,s budget as query length grows.}
\label{fig:latency}
\end{figure}

\begin{figure}[t]
\centering
\includegraphics[width=0.86\columnwidth]{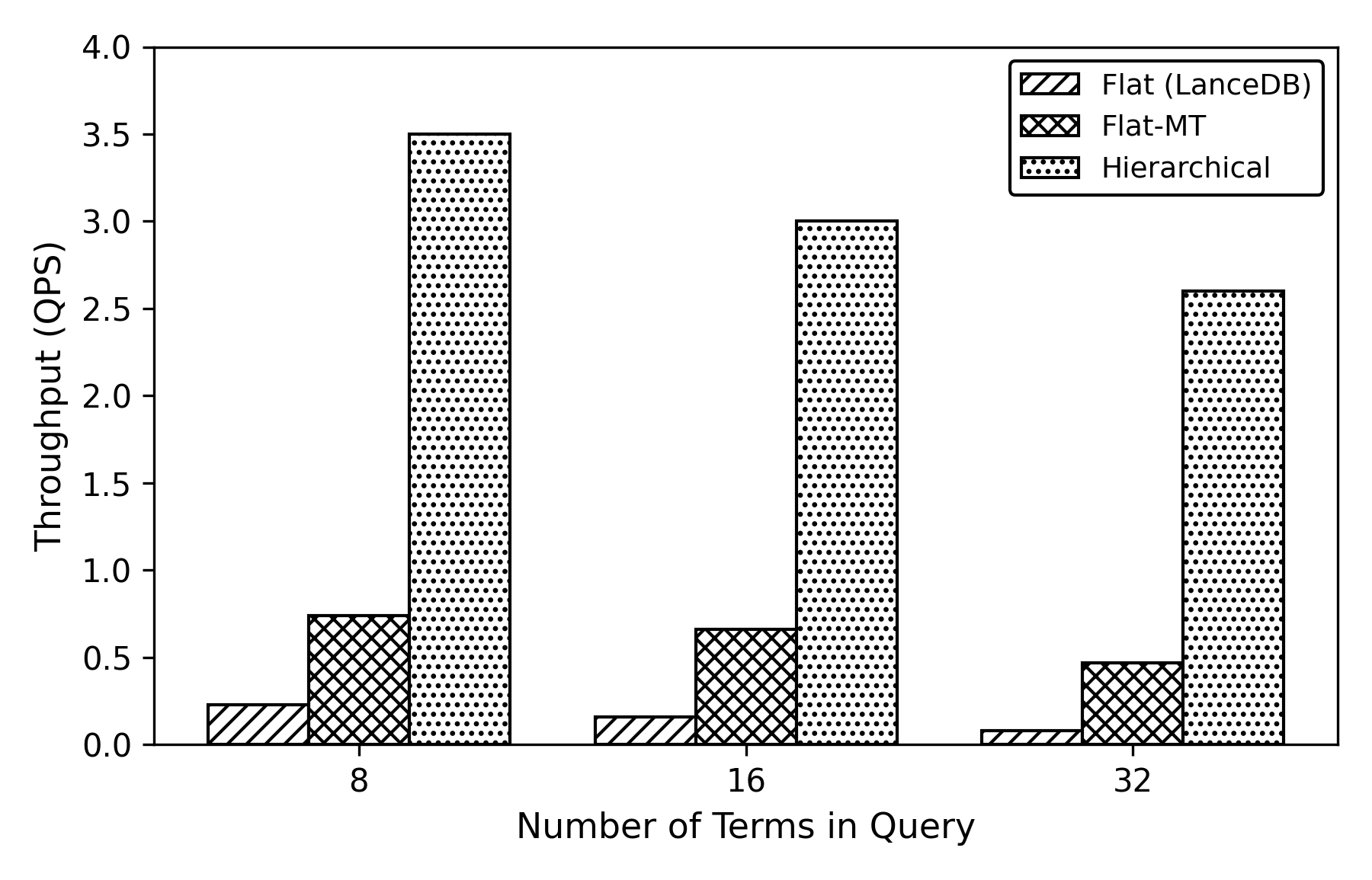}
\caption{Single-query throughput over 1B documents, by query length. This chart is derived, not separately measured: QPS $= 1000/\mathrm{ms}$ applied to the latencies of Figure~\ref{fig:latency}. Hierarchical BM25 holds 2.6--3.5 QPS while both flat baselines fall as queries lengthen: the gap vs.\ Flat-MT widens from 4.7$\times$ (8 terms) to 5.6$\times$ (32 terms), and vs.\ flat single-threaded from 15$\times$ to over 30$\times$.}
\label{fig:qps}
\end{figure}

\subsection{Single-query latency and throughput}
\label{sec:singleq}

Over one billion documents, Hierarchical BM25 answers 16-term queries in ${\sim}300$\,ms. It stays inside the one-second budget across 8-, 16-, and 32-term queries (Figure~\ref{fig:latency}). Against the flat multi-threaded baseline on the same hardware, it delivers 4.7--5.6$\times$ lower latency across the query-length range, with a resident footprint of ${\sim}4.4$\,GB versus ${\sim}400$\,GB (Table~\ref{tab:memory}). The gain is structural: the resident Level-1 filter removes most of the corpus before any disk-backed scoring, so the number of random NVMe reads per query drops sharply. The drop holds as query length---and therefore naive fan-out---grows from 8 to 32 terms, where the flat baselines degrade the most.

Latency and throughput are two views of the same single-query measurement. At one query in flight, throughput is simply the reciprocal of latency ($\mathrm{QPS} = 1000/\mathrm{ms}$). We plot both because each makes a different comparison easy. The latency plot (Figure~\ref{fig:latency}) makes the absolute one-second budget easy to check. The throughput plot (Figure~\ref{fig:qps}) makes the \emph{relative} gap between configurations readable directly as a multiplier.

Read this way, Hierarchical BM25 sustains 2.6--3.5 QPS single-threaded across 8--32 terms. That is 4.7--5.6$\times$ the flat multi-threaded baseline (0.47--0.74 QPS) and 15--31$\times$ the flat single-threaded baseline (0.08--0.23 QPS). The gap widens as query length grows. Hierarchical BM25's latency is nearly flat: 287--387\,ms, a $1.35\times$ increase from 8 to 32 terms. The flat baseline degrades far faster---$2.74\times$ over the same range---as postings fan-out increases. Flat-MT falls in between ($1.58\times$): threading parallelizes the scan but does not reduce how much of the corpus each query still has to touch.

This single-query throughput advantage is the baseline the parallel-query results in Section~\ref{sec:concurrent} build on. It already holds with no concurrency and no cache warming, before either of those further amplifies it. Note also that the widening gap with query length matches Section~\ref{sec:bmw}'s argument: Hierarchical BM25's cost is governed by a fixed cluster budget, not by how much the candidate pool grows as $q$ increases.

\begin{figure}[t]
\centering
\includegraphics[width=0.86\columnwidth]{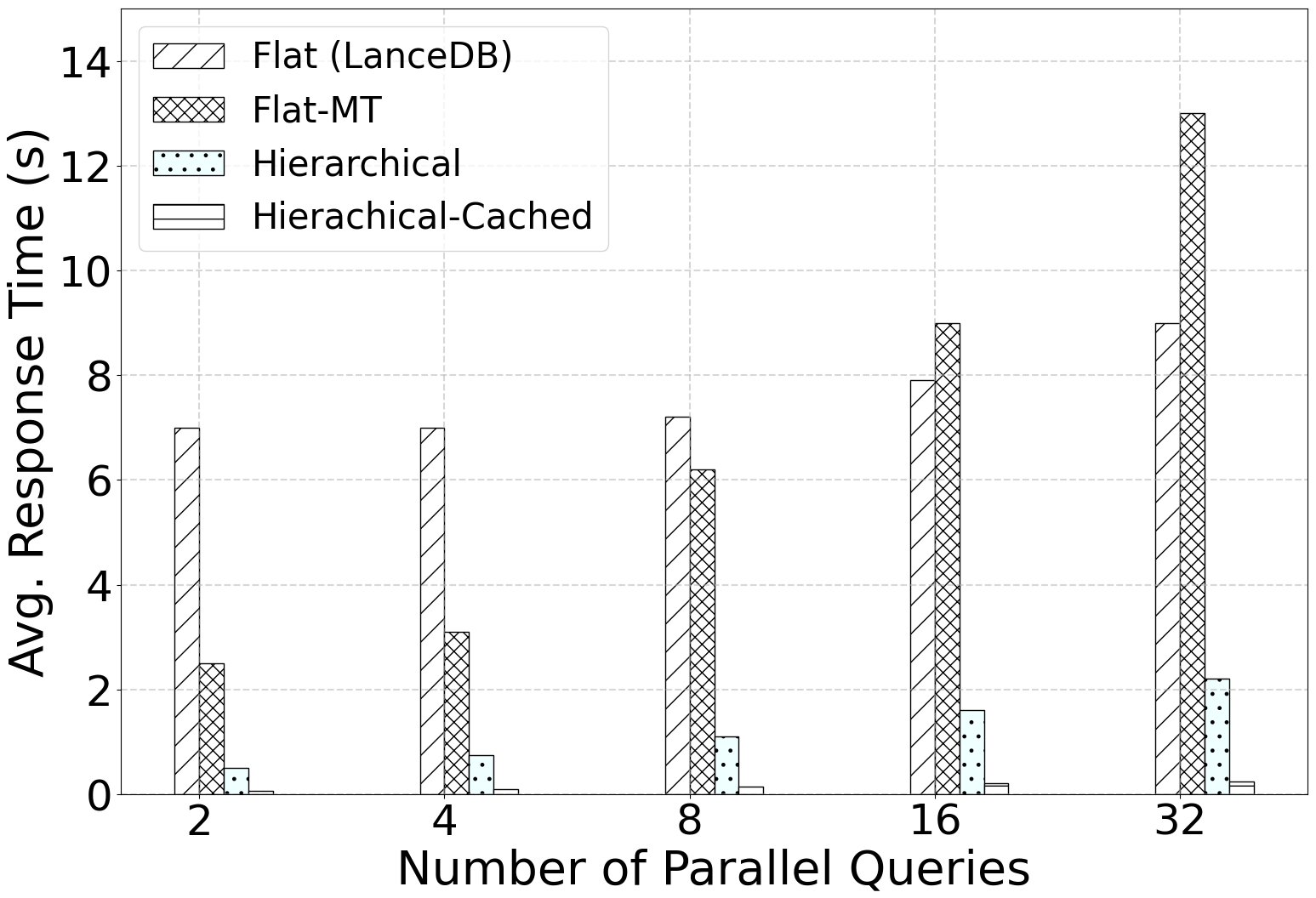}
\caption{Measured average response time under concurrent 16-term queries over 1B documents in 1K clusters (${\sim}4\%$ visited per hierarchical query). The flat baselines' latency grows with concurrency, because every query pays the full disk-bound fan-out and parallel queries contend for the same NVMe bandwidth; Hierarchical BM25 stays near its single-query latency until cores saturate.}
\label{fig:concurrent-latency}
\end{figure}

\begin{figure}[t]
\centering
\includegraphics[width=0.86\columnwidth]{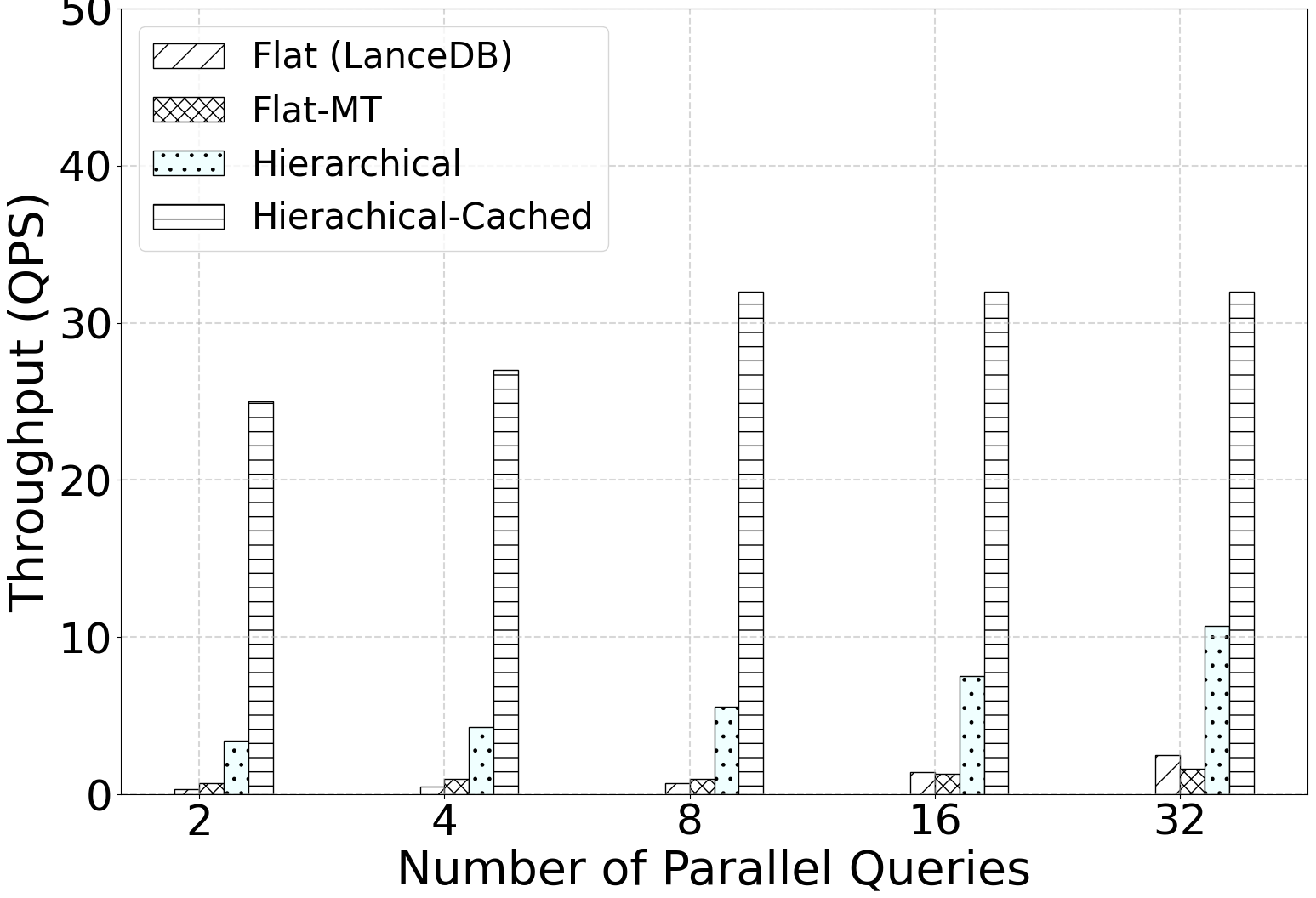}
\caption{Measured throughput under concurrent 16-term queries over 1B documents in 1K clusters (${\sim}4\%$ visited per hierarchical query). The flat baselines stay I/O-bound under 3\,QPS regardless of concurrency; Hierarchical BM25 scales with the number of parallel queries once Level-1 has pruned the disk-bound work, and a warmed Level-2 cache removes most of the remaining I/O cost.}
\label{fig:concurrent}
\end{figure}

\subsection{Throughput under concurrent queries}
\label{sec:concurrent}

Single-query latency bounds the tail, but a lexical index also has to sustain concurrent traffic. We hold the query mix fixed (16-term queries, as above) and issue 2, 4, 8, 16, and 32 queries in parallel against the same billion-document corpus and hardware. We compare four configurations: flat (LanceDB), flat multi-threaded, Hierarchical BM25, and Hierarchical BM25 with a warmed Level-2 cache (Hierarchical-Cached). The warmed configuration is the steady state once the fixed-size cache (Section~\ref{sec:hbm25}) has been populated by prior traffic, rather than measured cold.

Latency and throughput diverge sharply across the four configurations (Figures~\ref{fig:concurrent-latency} and~\ref{fig:concurrent}). On the latency side (Figure~\ref{fig:concurrent-latency}), the flat baselines' average response time climbs with concurrency---parallel queries contend for the same NVMe bandwidth while each still pays its full fan-out---whereas Hierarchical BM25 holds near its single-query latency. On the throughput side (Figure~\ref{fig:concurrent}), the flat baselines stay under 3 QPS even at 32 concurrent queries. Every query still pays the same disk-bound fan-out regardless of how many other queries run alongside it, and parallelism helps latency hiding only marginally when every query is I/O-bound on the same ${\sim}400$\,GB structure. Hierarchical BM25 (cold Level-2) reaches ${\sim}10.9$ QPS at 32 parallel queries---roughly 4--6$\times$ the flat multi-threaded baseline across the range, widening toward $6\times$ as concurrency increases. This is consistent with the Level-1 filter cutting the amount of disk-bound work per query, regardless of concurrency. Hierarchical BM25 with a warmed cache reaches ${\sim}$25--32 QPS from as few as 2 parallel queries and plateaus there. Once the frequently accessed Level-2 postings are cache-resident, throughput is limited by CPU scoring rather than NVMe I/O, so additional concurrency mostly fills otherwise-idle cores rather than contending for disk bandwidth.

\subsection{Selection quality against the exhaustive index}
\label{sec:quality}

\begin{figure}[t]
\centering
\includegraphics[width=0.86\columnwidth]{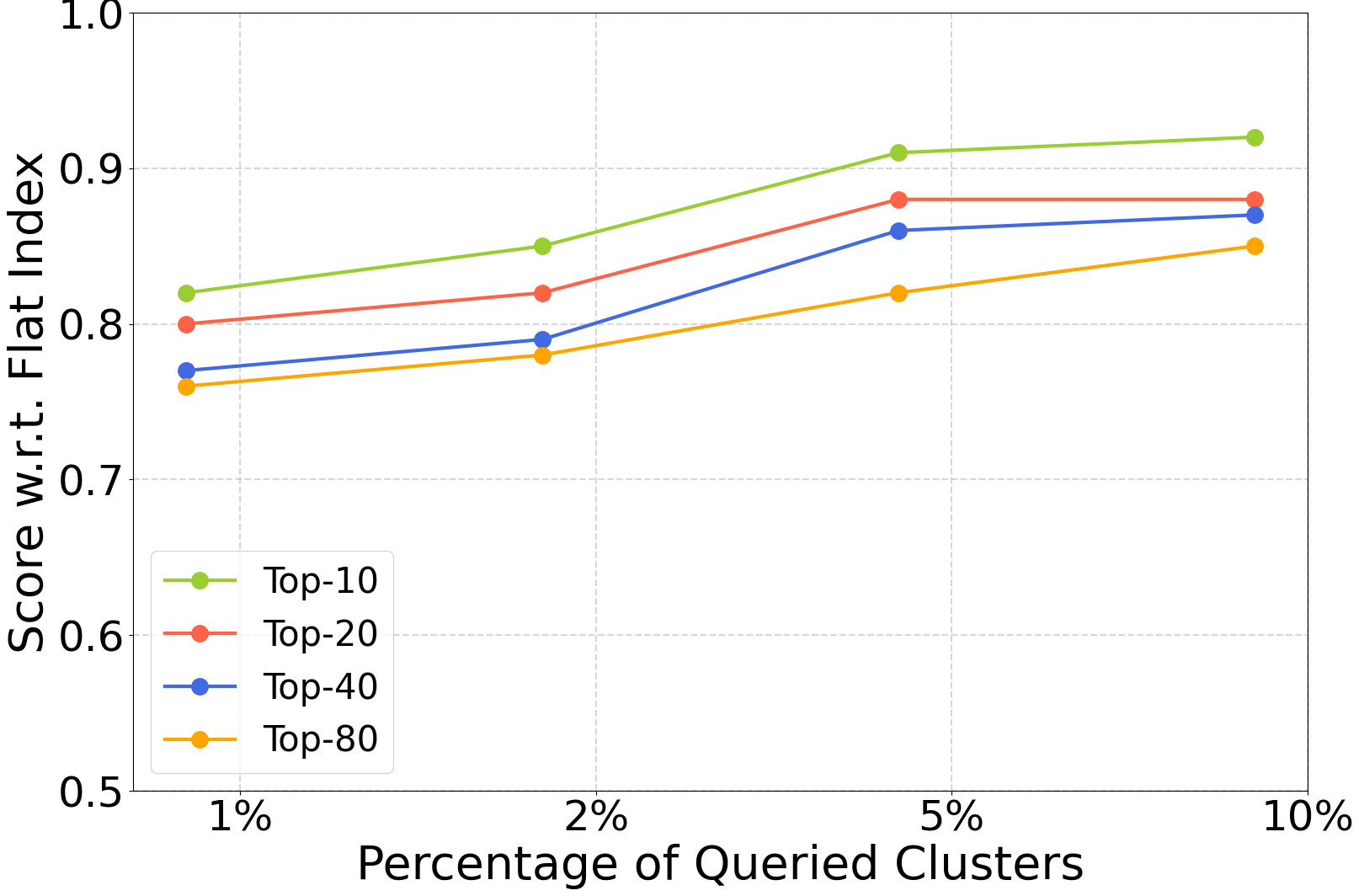}
\caption{Measured aggregate score of the hierarchical top-$k$ relative to the flat exhaustive index (1.0 = identical result strength), over 500K documents in 500 balanced clusters, sweeping the fraction of clusters visited and the result depth. Quality rises steeply with the visited fraction, and shallow result lists are preserved best.}
\label{fig:scoreratio}
\end{figure}

\begin{figure}[t]
\centering
\includegraphics[width=0.86\columnwidth]{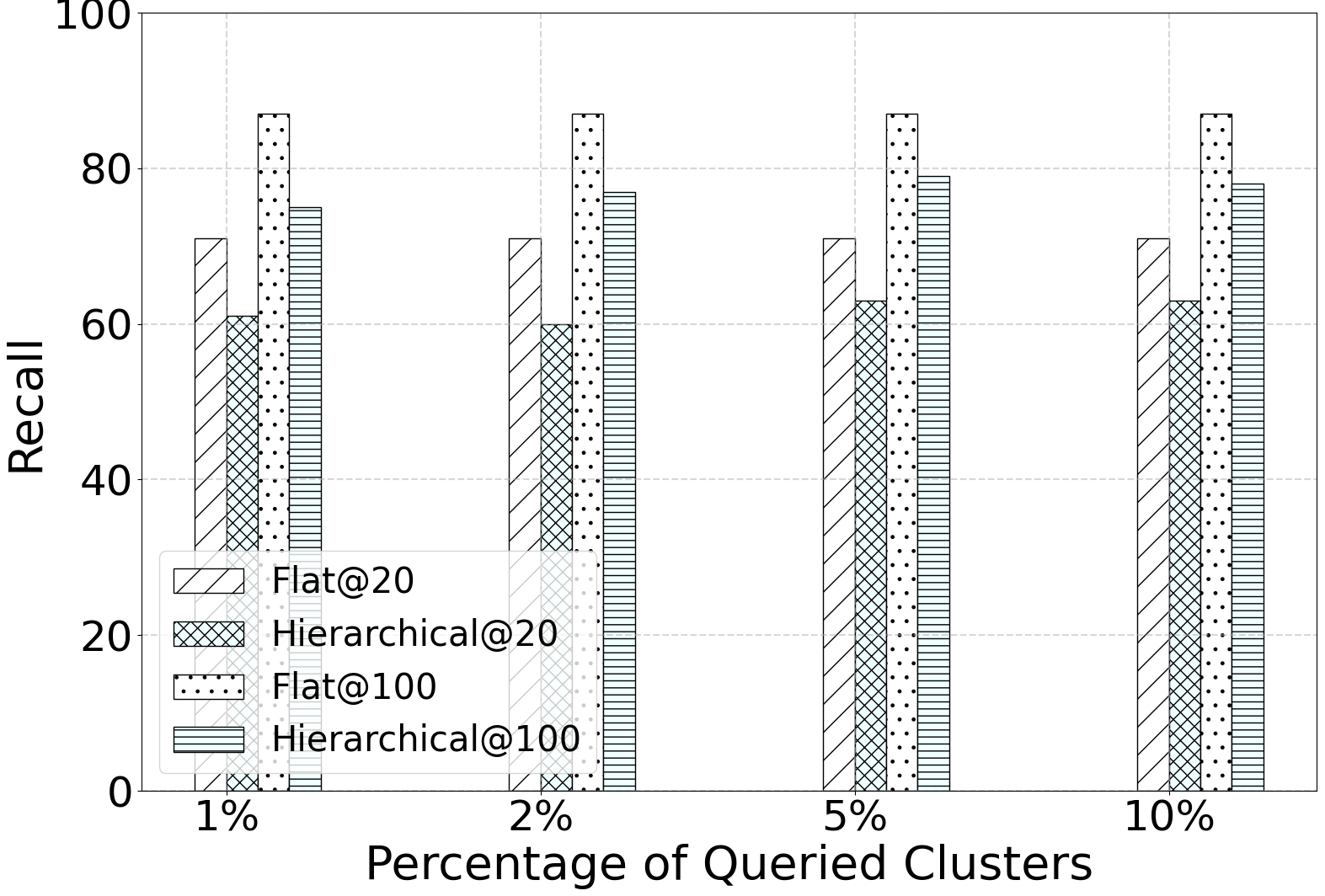}
\caption{Measured recall at depths 20 and 100 for hierarchical selection versus the flat exhaustive index over the same 500K-document configuration. Hierarchical recall climbs steeply as the visited fraction grows and approaches the flat index's own recall by 10\%.}
\label{fig:recall}
\end{figure}

The price side of the trade is measured at a smaller configuration: 500K documents in 500 balanced clusters, evaluated against the flat exhaustive index over the same corpus, sweeping the fraction of clusters visited per query from 1\% to 10\% and the result depth from top-10 to top-80. Two metrics capture two different failure modes. The first (Figure~\ref{fig:scoreratio}) is the aggregate score ratio: the summed BM25 score of the hierarchical top-$k$ divided by the flat index's, so 1.0 means the selected clusters contained results exactly as strong as exhaustive search found. The second (Figure~\ref{fig:recall}) is recall of the flat index's own result lists at depths 20 and 100.

Both move the same way. Visiting 1\% of clusters already recovers 0.76--0.83 of the exhaustive score, depending on depth; 5\% recovers 0.83--0.91; 10\% recovers 0.85--0.92. Shallow lists are preserved best (top-10 reaches 0.92 of the flat score at 10\% visited, top-80 reaches 0.85), and that is the shape the design predicts: the strongest documents concentrate in the strongest clusters, while a deeper list depends increasingly on middling documents scattered through clusters the selection skipped. Recall behaves the same: hierarchical recall at both depths climbs steeply from 1\% to 5\% visited and approaches the flat index's own recall by 10\%. The billion-document configuration of Sections~\ref{sec:singleq} and~\ref{sec:concurrent} visits ${\sim}4\%$ of clusters, which sits on the steep, favorable part of these curves.

Three caveats bound this measurement. It is taken at 500K documents and 500 clusters, not at $10^9$ and ${\sim}1$K: the trends match the design's argument, but billion-scale recall is extrapolated from them, not measured. Selection in this study is driven by the aggregate signal, because the query mix contains too few $T_M$ terms for the co-occurrence signal to fire (Section~\ref{sec:bcost}); so these curves measure the aggregate half of selection; isolating $B$'s contribution remains open. And the corpus shares the compact vocabulary of Section~\ref{sec:setup}, so quality on a natural-vocabulary corpus remains open as well.

\section{Discussion and Limitations}
\label{sec:discussion}

The most important limitation is that the price of the trade is only partly measured. Section~\ref{sec:trade} argued that giving up rank safety is worth structural bounds on memory and latency, on the grounds that a swap deep in the top-$k$ rarely changes a hybrid pipeline's output while a multi-second lexical query breaks it. That argument holds only if the approximation misses \emph{little}. Section~\ref{sec:quality} measures this directly at a 500K-document configuration: visiting 5--10\% of clusters recovers 0.83--0.92 of the exhaustive index's result score, and recall approaches the flat index's own by 10\%. This is consistent with the argument and covers the operating region the billion-document benchmark uses. What remains unmeasured is the same price at $10^9$ documents, on a natural-vocabulary corpus, and with the co-occurrence signal's contribution isolated; nDCG against relevance judgments is unmeasured everywhere. Until those measurements exist, the billion-scale latency and throughput results describe the cost side of a trade whose price is known only at smaller scale. Closing that gap remains the most important open step, ahead of even the BMW comparison in Section~\ref{sec:bmw}.

A second limitation concerns how the guarantee scales. It is structural for a fixed cluster count, but not for a growing corpus held at that fixed count. Holding ${\sim}1$K clusters fixed while the corpus grows increases the amount of data each visited cluster holds, and therefore the per-query work at a fixed top-$k_{\mathrm{clu}}$. Holding per-query work fixed by growing the cluster count instead grows Level-1's resident size---but that growth can be paid incrementally rather than through a global rebuild, by splitting only oversized clusters as they appear (Section~\ref{sec:clustering}). The ${\sim}4.4$\,GB figure and the sub-second guarantee both describe the one-billion-document operating point measured in Section~\ref{sec:eval}, not an asymptotic property independent of scale.

Related to this, adversarial or skewed query distributions degrade the guarantee, in a way that connects back to Section~\ref{sec:bmw}'s analysis. The cost bound depends on the Level-1 grouping being selective and the Level-2 cache achieving a high hit rate under the live query distribution. A workload that defeats grouping---for instance, one dominated by very common terms spread uniformly across clusters---pushes more work onto the disk-backed path. This failure mode is structural, not incidental: the terms such a workload leans on are exactly the high-df head that Section~\ref{sec:clustering} excludes from the clustering features \emph{because} they carry no topical signal, so no topical partition can concentrate them. By the same mechanism discussed in Section~\ref{sec:bmw}, such a workload would likely degrade WAND-family pruning at the same time, for a related reason. A further build-time consequence of Section~\ref{sec:clustering}: as a live corpus grows and its topic mix drifts, the balance enforced at build time erodes, so the partition needs maintenance flat indexes do not---incremental rebalancing by splitting oversized clusters (Section~\ref{sec:clustering}), which is local and keeps Level-2 scoring exactly correct, punctuated by occasional full reclustering to restore the global optimality that repeated splitting cannot.

Finally, this paper's scope is deliberately limited to the scaling problem alone. A separate line of work addresses lexical retrieval's blindness to the semantic vocabulary gap via query expansion. That method only rewrites the query text, and this paper's index answers ordinary BM25 queries regardless of how they were produced. The two therefore compose without architectural changes to either---though we have not measured the composition end-to-end.

\section{Conclusion}

At a billion documents, a flat BM25 index costs either ${\sim}400$\,GB of resident memory or disk-bound multi-second latency. Hierarchical BM25 trades rank safety for structural bounds on both. It bounds resident memory to ${\sim}4.4$\,GB with a resident coarse index plus a fixed-size fine cache. It holds single-query latency near 300\,ms---4.7--5.6$\times$ faster than a flat multi-threaded index---and sustains up to ${\sim}32$ QPS under concurrent load with a warmed cache, versus under 3 QPS for flat indexing. The sacrifice is confined to cluster selection. Every scored document receives its exact, corpus-wide BM25 score.

We identify and correct a cross-cluster scoring bug that silently biased rankings by cluster membership, fixed at a cost of ${\sim}100$\,KB of global statistics. We also argue, analytically, that Hierarchical BM25's cluster selection should outperform WAND-style dynamic pruning specifically at the long query lengths common in retrieval-augmented pipelines. It needs only single-term topical concentration, rather than the document-level multi-term co-occurrence WAND's pruning depends on. The aggregate signal alone dilutes as query length grows and cannot tell whether several terms' evidence came from one document or many; the design's second signal covers exactly that gap, tracking co-occurrence for the discriminative terms too spread out across clusters to self-select, at a cost governed by how many such terms a query contains rather than by query length.

All of this comes with the caveats stated throughout. Hierarchical BM25 is not rank-safe where BMW is. Selection quality is priced at a 500K-document configuration---0.83--0.92 of the exhaustive result score at 5--10\% of clusters visited---but not yet at billion scale, not on a natural vocabulary, and not with the co-occurrence signal's contribution isolated. Document-reordered BMW is the fair baseline still to be tested. Those measurements, and that comparison, are the open steps that would turn this architectural case into a settled one.

\balance

\section*{Acknowledgments}

We thank Cornel Constantinescu for his help in the investigation.

\end{document}